%Paper: hep-th/9312205
%From: rim@mips.chonbuk.ac.kr (rim chai ho )
%Date: Wed, 29 Dec 93 16:03:00 -0800

%\enput vanilla.sty
%\pagewidth{6.5 in}
%\pageheight{8.5 in}
%\TagsOnRight
%\scaletype{\magstep1}
%\baselineskip=28truept plus 0.2truept minus 0.2truept
%\parskip=10pt plus 1pt

\define \su{{su(1,1)}}
\define \suq {{su_{q^2}(1,1)}}
\define \N{{K_0}}
\define \qd{$q$-deformed }
\define \vac {{|vac\!\!>}}
\define \vaco {{|0\!\!>}}

\vskip
\centerline {\bf $q$-deformed oscillator  associated with Calogero model}
\centerline {\bf and its $q$-coherent  state }
\bigskip

\centerline
{K.\  H.\ Cho, Chaiho Rim and D.\ S.\ Soh}
\vskip -4pt
\centerline
{Department of Physics}
\vskip -8pt
\centerline
{Chonbuk National University}
\vskip -8pt
\centerline
{Chonju, 560-756, Korea}
\centerline
{S.\ U.\ Park}
\vskip -4pt
\centerline
{Department of Physics}
\vskip -8pt
\centerline
{Jeonju University}
\vskip -8pt
\centerline
{Chonju, 560-759, Korea}
\smallskip
\vfill

\centerline
{Abstract}
\smallskip
We present the $q$-deformed
para-bose oscillators associated with (two-body) Calogero model.
$q$-deformed coherent  state is also constructed
and its resolution of unity
is demonstrated.

\vfill\eject
\noindent
{\bf 1.  Introduction}
\bigskip

$q$-deformation [1] of the classical Lie algebra has been
an active research area recently.
For $q$-deformation of $\su$ with which
we are concerned in this paper, in literature
one can find its realizations
in terms of $q$-oscillators [2,11]  and $q$-parabose oscillators [3].
For \qd $osp(1,2n)$ algebra, we refer to [4].

Very recently, it is demonstrated that
there is an oscillator analogy [5]
for Calogero model [6]
although differential operator realization
[7] of $\su$ is known for a long time.
The newly appeared oscillator
is not the ordinary harmonic oscillator
in the sense that it contains  exchange operators in the commutation
relations.
Therefore, it  might be natural to study the $q$-deformation
of the  modified oscillator, which will
expose  different behaviour from that of the ordinary oscillators.

We will confine ourselves to the two-body Calogero
model since this is the non-trivial simplest case.
The modified oscillator arises from the relative motion part,
which contains the inverse square interation.
The modified oscillator is, however, not the new one:
This oscillator realises para-bose algebra [8] and becomes
a  para-bose oscillator.
This  kind of oscillator
has  already been studied [9] with the introduction of parity
operator (exchange operator in two-body case).
It is also known that para-bose algebra is in general
isomorphic to $osp(1,2n)$ super-algebra [10].
On the other hand, $q$-deformation of this oscillator system is not
thoroughly studied  as far as we are aware of.
In this paper, we will present the $q$-oscillator
realization such that it encompasses the Calogero model
and the para-bose oscillator as well
and construct its coherent state with resolution of unity.

This paper is organized as follows. In Section 2, we
give a brief review on the $q$-deformation of $\su$, \,\,
and we give a realization of  the $\suq$ generators associated with
(two-body) Calogero
model in terms of the \qd modified oscillators. The realization
looks same as that  in the standard $q$-oscillators [11]
except the appearance of the exchange operator (see Eqs. (26-28)).
This \qd modified oscillators provide an explicit operator realization of
$q$-parabose or $osp_{q^2}(1,2)$ super-algebra.
A slightly different $q$-deformation which covers the modified oscillator has
also been considered recently in Ref. [12].
In Section 3, we construct the $q$-deformed  coherent  state  for
the  para-bose oscillator. We demonstrate the resolution of unity for the
\qd coherent state following the method of Ref. [13] given for the
undeformed para-bose coherent state.
This  is  the  generalization  of  that  of  $q$-deformed
$osp(1,2) $ superalgebra considered in Ref. [14].
In Section 4,
summary of our results and some remarks are given.
\bigskip
\centerline{\bf 2.  $q$-deformation of para-bose oscillator }
\bigskip

The $ su(1,1)$ satisfies the commutation relations
$$
[K_0 , K_{\pm}]  =  \pm K_{\pm} \,, \qquad
[K_+ , K_-]  =  -2K_0\,,
\tag 1
$$
and  Casimir operator is given as
$$
C = K_0 (K_0 -1) - K_+  K_-\, .
\tag 2
$$
Its $q$-deformation, $su_{q^2}(1,1)$, is given [2] as
$$
[\tilde K_0, \tilde K_{\pm}] = \pm  \tilde K_{\pm}\, , \qquad
[\tilde K_+, \tilde K_-] = -[2 \tilde K_0]_{q^2}\, ,
\tag 3
$$
where $[x]_{\mu} \equiv (\mu^x - \mu^{-x}) / (\mu - \mu^{-1})$.

To find a realization of the $\suq$ in terms of the
generators of $\su$ we make an ansatz [15]:
$$
\tilde K_0 =  K_0 \, , \qquad
\tilde K_+ = \frac 2 {[2]_q}  F(K_0) K_+  \, ,\qquad
\tilde K_- = \frac 2 {[2]_q}   K_- F(\N)\, .
\tag 4
$$
Then to satisfy the the $q$-deformed  algebra Eq.\ (3),
we have a recursion relation
$$
(\frac {2}{[2]_{q^2}})^2 (G(\N)- G(\N +1)) = - [2\N ]_{q^2}\, ,
\tag 5
$$
where
$$
G(\N) \equiv F^2(\N) K_+ K_-
= F^2(\N) ( ( \N (\N -1) - C),
\tag 6
$$
and $C$ is the Casimir operator given in Eq.\ (2).

To find $G(\N)$, we first note that $F(\N)$ in Eq.\ (4)
should be a non-singular operator on the Hilbert space.
Therefore, for the vacuum defined as $K_- \vac = 0$ and
$\N \vac = k_0 \vac$, we have
$$ G(\N = k_0) = 0
\tag 7
$$
to make the operator consistent. From the recursion relation
Eq.\ (5) and the initial condition Eq.\ (7), we have
$
G(\N) = \frac 14 [2\N -2 k_0]_q [2\N + 2k_0 -2]_q$
and this leads to the  $F(\N)$ as
$$
F(\N) = \sqrt {\frac {[2\N- 2k_0]_q [2 \N + 2k_0-2]_q }
{(2\N -2k_0) (2\N + 2k_0-2) }}\,.
\tag 8
$$
In this case the \qd Casimir operator is  given as
$$
\tilde C =[\tilde \N]_{q^2} [\tilde \N -1]_{q^2}
             -\tilde K_+ \tilde K_-
         = [k_0]_{q^2} [k_0-1]_{q^2} \,.
\tag 9
$$

Let us apply this formalism to the
ordinary harmonic oscillator
system.
This system is described by
the creation and annihilation operators
$a = (x+ip)/ \sqrt 2$, $a^+  = (x-ip) / \sqrt 2$.
$\su$ generators are given as
$\N =\frac 12 (n+\frac 12)$,
$K_+ = \frac 12 (a^+)^2$ and $K_- =\frac 12 a^2$
where $n =a^+a$.
Casimir operator is given as
$C= -3/16$.
This system has two different representations
whose $k_0$ is 1/4 and
3/4 respectively.
We find that $F(\N)$ in Eq.\ (8) is same for both cases:
$$
F(\N) = \sqrt {\frac {[2\N- \frac 12]_q [ 2\N -\frac 32]_q }
{(2\N -\frac 12) (2\N -\frac 32) }}\,.
\tag 10
$$
Therefore, the generators of $\suq$ are given as
$$
\tilde K_0 = K_0 \,, \quad
\tilde K_{+} = \frac{2}{[2]_q}
               \sqrt {\frac {[n]_q [n-1]_q }{n (n-1) }} K_+\,, \quad
\tilde K_{-} = \frac{2}{[2]_q} K_-
               \sqrt {\frac {[n]_q [n-1]_q }{n (n-1) }}\,.
\tag 11
$$
The \qd \,Casimir operator  is given as
$\tilde C
= -[\frac 14]_{q^2} [\frac 34]_{q^2} $.
One can realize the   algebra $su_{q^2}(1,1)$ in terms of the
$q$-deformed oscillator [2,11]. From  Eq.\ (11), we get
$$
\tilde K_0 = \frac 12 (a^+ a + \frac 12)\,, \qquad
\tilde K_+ = \frac 1{[2]_q} (a_q^+)^2 \,,\qquad
\tilde K_- = \frac 1{[2]_q}   a_q^2\,,
\tag 12
$$
where
$$
a_q = a \sqrt {\frac {[n]_q}{n} }, \qquad
a_q^+ = \sqrt {\frac {[n]_q}{n}} a^+ \,.
\tag 13
$$

For other represenatation of $\su$ we consider
the relative motion part of the two particle
Calogero model.
The two particle Calogero model is given as
$$
H= \frac 1{2m} (p_1^2 + p_2^2)
   + \frac {m\omega^2}{16} (x_1-x_2)^2 + \frac g{(x_1 - x_2)^2}\, .
\tag 14a
$$
If we rescale $x_i \to \rho x_i$ and
$p_i \to \frac \hbar \rho p_i$
where $\rho = \sqrt {\frac {\hbar }{m\omega}}$,
then we have the hamiltonian
$$
H= \hbar\omega [\frac 12 (p_1^2 + p_2^2) + \frac1{16} (x_1-x_2)^2
+ \frac \lambda {(x_1 - x_2)^2}]\,,
\tag 14b
$$
where $ \lambda = \frac {mg}{\hbar^2} \ge -1/4$.
Restricting to the relative motion part only and
denoting $x_1 - x_2 =2x$ and $p_1 -p_2 =p$ such that $[x, p] =i$,
we have in the unit of $ \hbar\omega $
$$
K_0  =  \frac 14 (p^2 + x^2+  \frac {\lambda}{x^2})\,,
\tag 15
$$
The other $\su$ generators are given as [7]
$$
K_{\pm}  =\frac 14 (-p^2 + x^2 \mp i (xp + px)
           -\frac {\lambda}{x^2}) \,.
\tag 16
$$
The Casimir operator has the value $\frac{\lambda}4 - \frac 3{16}$,
which differs from that of the harmonic oscillator system
($\lambda =0$). Now  \qd \,generators are obtained if we put
$$
F(\N) = \sqrt {\frac {[2\N- 1 - \sqrt{\frac 14 + \lambda}]_q
[ 2\N -1 + \sqrt{\frac 14 + \lambda}]_q }
{(2\N -1 - \sqrt{\frac 14 + \lambda})
(2\N -1 + \sqrt{\frac 14 + \lambda}) }}\,.
\tag 17
$$

For the Calogero model it is also possible to realize
$\su$ in terms of a modified oscillator system by exploiting
an exchange operator formalism [5]. We will present the
\qd \,version of the system paying attention to the role of the
exchange operator.

Let us  introduce notations.
$$
\pi = p + i \frac lx \,\, M,
\tag 18
$$
where $l$ is a real parameter to be determined in terms of
$\lambda$ in the Calogero model.  $M$ is an exchange
operator in the two-body system and plays the role of the
parity operator in this reduced one-body system,  {\it i.e.},
$$
Mp = -pM \, , \qquad  Mx = -xM \, , \qquad  M = M^+ = M^{-1}\, .
\tag 19
$$

Next define ladder operators $A$ and $A^+$ as
$$
A = \frac 1{\sqrt 2} (x + i\pi)\, ,\qquad
A^{+} = \frac 1{\sqrt 2} (x - i\pi)  \, .
\tag 20
$$
Then  the modified hamiltonian is defined as
$$
K_0 \equiv \frac 14 (A^+ A + A A^+ )
=\frac 14 (p^2 + x^2 +
 \frac {l(l-M)}{x^2})  \,.
\tag 21
$$

One may realize $su(1,1)$ algebra just like in the harmonic
oscillator case by identifying
$$
K_0 = \frac 12 ({ N } + \frac 12)\,, \qquad
K_+ = \frac 12 (A^{+})^{2} \,, \qquad
K_- = \frac 12 A^{2}\,.
\tag 22
$$
Like in harmonic oscillator system, we define $\N$ in terms of $N$.
The Casimir operator is given as $ { C}= -3/16 +l(l-M)/4$.
One may easily  check the algebraic relations
if one uses  commutation relations
$$
[{ N}, A^{+}] = A^{+}\, , \qquad
[{ N}, A] = -A \, , \qquad
[A, A^+] = 1 + 2lM  \, ,
\tag 23
$$
which are obtained from their definitions.

The Hilbert space is obtained by applying the ladder operator
${ K}_+$ successively on the vacuum state
which is annihilated by ${ K}_-$.
On the other hand, the parity operator $ M$  commutes
with ${ K_0}$ and ${ K}_{\pm}$.
Therefore, the
$M$ behaves like a number as far as this $su(1,1)$ is concerned.
One can obtain the eigenstates of the two-boson (two-fermion)
Calogero system by restricting the Hilbert space of ${ K}_0$ to
a subspace of symmetric (anti-symmetric) eigenstates with $\lambda$
identified as $l(l-1)$ ($ l(l+1)$).

To realize the  generators of $\suq$ in Eq.\ (4) in terms of
the  $q$-deformed modified oscillator, we note first that
$k_0$ for one representation can be obtained from the vacuum
$|0>$ defined by $A|0> =0$, which gives $k_0 = 1/4 + lM/2$.
The vacuum for the  other reprentation, $ A^+ |0>$ gives
$k_0 = 3/4 - lM/2$.
Referring to Eq.\ (8), we have
$$
F({\N}) = \sqrt {\frac {[{N}-lM]_q [{ N}+lM-1]_q}
{({ N}-lM) ({ N}+lM-1)}},
\tag 24
$$
independent of the representations.
$k_0$ is no longer a number
when the creation and annihilation operator $A^+$, $A$ are present.
The Casimir operator is given as
$$
\tilde { C} =  [\frac 14 + \frac
{lM}{2}]_{q^2} [-\frac 34 + \frac {lM}{2}]_{q^2}\,.
\tag 25
$$

Now we define  the $q$-deformed generators in Eq.\ (4) in terms of
\qd oscillators as
$$
\tilde K_0  \equiv K_0\,,\qquad
\tilde { K}_+ \equiv \frac 1{[2]_q} (A^+_q)^2\, , \qquad
\tilde { K}_- \equiv \frac 1{[2]_q} (A_q)^2\, .
\tag 26
$$
Then Eqs.\ (22) and (24) give
$$
A_q  =  A \sqrt {\frac {[{ N}-lM]_q}{{ N}-lM}} \, , \qquad
A^{+}_q  =  \sqrt {\frac {[{ N}-lM]_q}{{ N}-lM}} A^{+}\, .
\tag 27
$$
They satisfy the relations
$$
\align
[{ N}, A^{+}_q] = A^{+}_q \, , \qquad
                       &[{N}, A_q] = -A_q\, , \\
A^{+}_q A_q  = [{N} - lM]_q\, , \quad
                       & A_q A^{+}_q  =[{N} + 1 +lM]_q\, .
\tag 28
\endalign
$$
\qd operators given in Eqs. (26-27) constitute a
representation of  \qd $osp(1,2)$
super-algebra.
One may use this explicit form to calculate Casimir operator of
$osp_{q^2}(1,2)$
$$
\align
 \tilde {C} &= [{ K}_0]_{q^2} [{ K}_0-1]_{q^2}
        -\tilde { K}_+ \tilde{ K}_-
        + [\frac 12]_{q^2} [\frac 14]_{q^2}
        \frac {\cosh \eta l}{\cosh 2 \eta { K}_0}[A_q, A_q^+]\\
        &= [\frac 14]_{q^2} (2[\frac 14]_{q^2} - [\frac 34]_{q^2})
        \cosh ^2 \eta l
        + \frac 14 ([2]_q + [2]_{q^2})( [\frac l 2]_{q^2})^2\,,
\tag 29
\endalign
$$
where $q= e^\eta$ and Eq.\ (29)
reduces to $-\frac 1 {16} + \frac {l^2}{4} $ when $q=1$.

\bigskip
\centerline {\bf 3. $q$-deformed coherent states}
\bigskip

In this section we will construct  the coherent states
for  the  $q$-deformed  modified  oscillator
obtained above and demonstrate the resolution of unity.
Let  us  begin  by  summarizing  the  Fock  space  representation  of   the
$q$-parabose
system.  Recalling that the Fock space  is unchanged under the
$q$-deformation,
we label them as $| n >, \quad n = 0, 1, 2, \cdots$.
The vacuum $\vaco$ is defined by
$$
A  \vaco = 0,  \qquad  M \vaco = \vaco ,
\tag 30
$$
and therefore from Eqs. (21)-(23),
$$
K_0 \vaco = \beta \vaco, \quad \beta \equiv \frac 14 + \frac l2, \quad
N \vaco = l \vaco.
\tag 31
$$
Since $K_0$ is positive-definite, it is clear that $l \geq -  \frac12  $.
The
ortho-nomalized ket $| n >$  is
given as
$$
| n \!\! > = \frac {1}{\sqrt {C_n !}}
A^{+n} \vaco = \frac {1}{\sqrt {\tilde C_n!}}
A^{+n}_q \vaco,
\tag 32
$$
where
$$
C_n = n + l (1 - (-1)^n),\quad \tilde C_n = [C_n]_q,
\tag 33
$$
and we use a notation  $f_n ! \equiv \Pi^n_{i=1} f_i$, and $0! = 1$.
We also note for later convenience,
$$
A^+_q |n \!\!> = \sqrt{\tilde C_{n+1} } |n + 1\!\! >, \quad
                A_q |n\!\! > = \sqrt {\tilde C_n} |n - 1 \!\!>,
\tag 34
$$
$$
N |n \!\!> = (n + l) | n\!\! >,\quad  M |n\!\! > = (-1)^n |n \!\!>.
\tag 35
$$

We recall that the set $\{ |n\!\! >\}$ forms a single irreducible
representation  of $osp(1,2)$ algebra with Casimir
$C (osp (1,2)) = - \frac 1{16} + \frac {l^2}{4}.$   On the
other hand, $\{|2p \!\!>\}$ and $\{|2p+1 \!\!>\}$ for
$ p  =  0,  1,  2,  \cdots $
form    two   distinct, parity-even  and  odd   irreducible   representations
$D_{\beta}$ and $D_{\beta + \frac12}$ of $su (1,1)$ \, algebra respectively.
We will label the eigenstates as $|k,p \!> :$
$$
|k, p \!> = \left\{\matrix |2p \!>  &  k=\beta \\
                |2p+1 \!\!>  & k=\beta+\frac12     \endmatrix \right\},
\qquad
K_0 |k, p\! > = (k+p)|k, p \!>.
\tag 36
$$
It is simple to show from Eqs. (22), (26), (32)-(36),
$$
\align
|k, p \!> &= \frac{1}{\sqrt{d_p!}} K^p_+ |k, 0 \!\!> =
\frac {1}{\sqrt {\tilde d_p !}} \tilde K_+^p |k, 0 \!\!>, \\
\tilde K_+ |k, p \!> &= \sqrt {\tilde d_{p+1}} |k, p+1 \!\!>, \quad
\tilde K_- |k, p \!> = \sqrt {\tilde d_p} |k, p-1\!\!>\,.
\tag 37
\endalign
$$
where
$$
d_p  = p (p + 2k-1), \qquad  \tilde d_p = [p]_{q^2} [p+2k-1]_{q^2}.
\tag 38
$$

We will use the definition of the unnormalized $q$-coherent state $|z)$ for
the
$q$-parabose oscillator as
$$
A_q |z) = z|z),
\tag 39
$$
where $z$ is a complex number.  Then $|z)$ is given by
$$
|z)  \equiv \sum^{\infty}_{n=0}
                \frac {z^n}{\sqrt{\tilde C_n !}}
                |n \!\! > = \left\{\sum^{\infty}_{n=0}
                \frac{z^n}{\tilde C_n !} (A^+_q)^n \right\} |0 \!\!>.
\tag 40
$$
We can associate with each normalizable ket $|\phi >,$
$$
| \phi > = \sum^{\infty}_{n=0}  \,\,\,\, |n ><n | \phi > \equiv
\sum^{\infty}_{n=0}
        \phi_n |n >,
\tag 41
$$
an entire function $\phi(z)$ defind as
$$
\phi (z) \,\,\,\, \equiv \,\,\,\, (z^* |\phi > = \sum^{\infty}_{n=0}\,\,\,\,
                \frac {\phi_n}{\sqrt{\widetilde C_n !}} z^n.
\tag 42
$$
Combining Eqs. (39) and (42), we get
$$
(z^* |A_q^+| \phi > = z (z^* | \phi > = z \phi (z),
\tag 43
$$
which shows that $A^+_q$ acts as a multiplication by $z$ in the
Bargmann space.  Taking $|\phi >$  as  $A_q  |\psi  >$  in  Eq.(43),  and
using
Eq.(28),
$$
z (z^* |A_q|\psi > = (z^* |A_q^+ A_q | \psi > = (z^* |[N - lM]_q| \psi >.
\tag 44
$$
Using Eqs. (33), (35), and (42), Eq.(44) shows that $A_q$ acts  as  an
operator
$(q=e^{\eta})$
$$
A_q \to (\cosh^2 \eta l + M \sinh^2 \eta l) \frac {d}{d(z;q)}
        + \frac {[2l]_q}{4z} (T_q + T_q^{-1})(1-M).
\tag 45
$$
Here, the parity operator $M$ acts as
$$
M \psi (z) = \psi (-z),
\tag 46
$$
and the $q$-derivative and $q$-shift operators are given as usual by
$$
T_q \psi(z) = \psi (qz),
\tag 47
$$
$$
\frac  {d}{d(z;q)}  \psi  (z)   =  \,\,  \frac    1z\,\,\,    \frac    {T_q
-
T_q^{-1}}{(q-q^{-1})}
\psi(z).
\tag 48
$$

We need a resolution of unity to complete the $q$-coherent state
descriptions.   It will provide a natural inner product for the Bargmann
space,
under which $A_q$ and $A^+_q$ are hermitian conjugates mutually.   To this
end,
we proceed in  an  analogous  way  taken   in   Ref. [13]   for   the
para-bose  coherent states. The essence of the method lies in finding the
correct measure of $|z)$ coherent states by using those of the $\su$
coherent states.

Let  us  introduce  the  $su_{q^2}$(1,1)
coherent  states   as    in  Refs. [13,16]
(For further discussions on $su_q$(2), $su_q$(1,1) coherent states, see [17]) :
$$
| \omega ; k )  \equiv  \sum^{\infty}_{p=0} \frac {1}{\sqrt {\widetilde d_p !}}
        \omega^p |k, p > \,\,\,
        = \,\,\,{\sum^{\infty}_{p=0} \frac {1}{\widetilde d_p !}
          \omega^p \widetilde K^p_+ } |k, 0>,
\tag 49
$$
where  $\omega$  is  a  complex  number  and  other  notations  are   given
in
Eqs.(36)-(38) and satisfy
$$
\tilde K_- |\omega ; k) = \omega |\omega ; k)\,.
\tag 50
$$

For a general ket $|g>$ in $D_k$,
$$
|g> = \sum^{\infty}_{p=0} | k, p > < k, p | g > = \sum^{\infty}_{p=0}
        g_p | k, p >,
\tag 51
$$
we may associate a complex function $g(\omega)$ defined as
$$
g(\omega) \equiv (\omega^* ; k|g> = \sum^{\infty}_{p=0}
        \frac{g_p}{\sqrt{\widetilde d_p!}} \omega^p.
\tag 52
$$
The action of $su_{q^2} (1,1)$ generators on
$g(\omega)$ is represented as
$$
 K_0  \rightarrow  k  +  \omega
\frac d{d\omega},
\tag 53
$$
$$
\tilde {K}_+ \rightarrow \omega, \qquad
\tilde {K}_- \rightarrow \left\{ [2k]_{q^2} \frac {T_{q^2}+ T_{q^2}^{-1}}2
        + \omega \cosh 4 \eta k \frac {d}{d(\omega;q^2)} \right\}
        \frac {d}{d(\omega;q^2)}.
\tag 54
$$

The $q$-coherent state $|z)$ in Eq.(40) can be  written in terms of
$|\omega ; \beta)$  and   $|\omega ; \beta+\frac12)$
compactly
as
$$
|z) = |\omega;\beta) + \frac{z}{\sqrt{[4\beta]_q}} | \omega;\beta+\frac12),
\tag 55
$$
with $\omega $ being identified as
$$
\omega = \frac{z^2}{[2]_q},
\tag 56
$$
with help of Eqs.(32)-(38).
Therefore, $\phi (z)$ in Eq.(42) is  decomposed
into even and odd parts:
$$
\align
\phi (z) &= \phi_+ (z) + \phi_- (z),\\
\phi_+ (z) &= (\omega^* ; \beta | \phi > \equiv \phi_1 (\omega),\\
\phi_- (z) &= \frac {z}{\sqrt{[4\beta]_q}} (\omega^* ; \beta + \frac12 | \phi >
                \equiv \frac {z}{\sqrt{[4\beta]_q}} \phi_2 (\omega).
\tag 57
\endalign
$$

We remark in passing that if we express $\phi (z)$ as a column vector
$$
\left(\matrix \phi_+ (z) \\ \phi_- (z) \endmatrix \right),
$$
then the operators are realized by matrix operators as  follows
(remind  that $A_q$ and $A_q^+$ change the parity):
$$
A^+_q \rightarrow \left(\matrix 0 & z \\ z & 0 \endmatrix \right),
\tag 58
$$
$$
A_q \rightarrow
\left(\matrix 0 &
\frac12(q^{2l}+q^{-2l}) \frac d{d(z;q)} + \frac {[2l]_q}{2z} (T_q + T^{-1}_q)\\
\frac d{d(z;q)} & 0 \endmatrix \right)\,.
\tag 59
$$
$\tilde K_{\pm}$ and $K_0$ are given as a  diagonal matrix in this basis and
their explicit form can be obtained using the realtions in Eq. (26).
A similar consideration has been done  in Ref. [18] especially
in connection with $q$-special functions.

To demonstate the resolution of unity for $|z)$,
suppose we have obtained a resolution of unity for $| \omega ; k) (k =
\beta$
or $\beta + \frac 12)$ coherent states in Eq.(49), namely
$$
\int d^2 (\omega ; q^2) G_k (|\omega|) \,\, |\omega ; k )
        (\omega ; k | = \sum^{\infty}_{p=0} |k, p ><k, p|=I,
\tag 60
$$
where
$$
d^2 (\omega ; q^2) \equiv \frac 12 d(|\omega|^2 ; q^2) d \theta
                = \frac {[2]_q}{2} |\omega| d (|\omega| ; q) d \theta,
\tag 61
$$
and $d (|\omega| ; q)$ is a standard $q$-integration [19, 20] and $d  \theta$
is
an
ordinary integration from 0 to $2 \pi$.
Then, using the definition of an inner product  in $D_k$
$$
<\! g'|g \!> = \sum^{\infty}_{p=0} g_p'^{*} g_p = \int d^2 (\omega ; q^2) G_k
(|\omega|)
        g'^{*}(\omega) g(\omega),
\tag 62
$$
where $g (\omega), g'(\omega)$ are associated with $|g \!>$,
$ |g' \! >$ as in Eqs.(51-52),
we have
$$
<\! \phi' | \phi \!> = \int d^2 (\omega ; q^2) \left\{G_{\beta} (|\omega|)
\phi_1'^{*} (\omega)
        \phi_1 (\omega) + G_{{\beta} + \frac 12} (|\omega|) \phi_2'^{*}
(\omega)
        \phi_2 (\omega)\right\},
\tag 63
$$
for arbitrary two kets $| \phi >$ and $| \phi' >$ in the  total  Fock  space
by
referring to Eqs.(36) and (57). In Eq.(63), we make the change of variable  as
in Eq.(56) and take into account that $\omega$ covers the  complex  plane
twice
while $z$ covers once :
$$
d^2 (\omega ; q^2) \rightarrow  \frac {|z|^2}{[2]_q} d^2 (z ; q).
\tag 64
$$
Collecting Eqs.(57), (63) and  (64)  and  observing  $|\phi  >,  |\phi'  >$
are
arbitrary, we end up with the resolution of unity for $|z)$ coherent states :
$$
\align
\frac 1{2[2]_q} \int d^2 (z ; q)
        & \left[ \left\{ |z|^2 G_{\beta} \left(\frac{|z|^2}{[2]_q}\right)
        + [4\beta]_q G_{\beta + \frac12} \left(\frac{|z|^2}{[2]_q}\right)
        \right\} |z)(z| \right.\\
        & + \left. \left\{ |z|^2 G_{\beta} \left(\frac{|z|^2}{[2]_q}\right)
        - [4\beta]_q G_{\beta + \frac12} \left(\frac{|z|^2}{[2]_q}\right)
        \right\} |z)(-z| \right] = I.
\tag 65
\endalign
$$
Thus, the problem to prove the resolution of unity for $|z)$ coherent states
reduces to finding  $G_k (|\omega|)$ satisfying Eq.(60).

The diagonal element of Eq. (60) in $|k, p >$ basis gives
$$
\int d(u ; q) u^{2p+1} G_k(u) = \frac 1{\pi [2]_q }
\prod _{j=1}^p ([j]_{q^2 }  [j+2k-1]_{q^2})\,,
\tag 66
$$
where we have put $|\omega| = u$, and $\theta$-integration is done.
To find $G_k(u)$,  we  recall the undeformed one,
$G_k^{(0)} (u)$ [13],
$$
G_k^{(0)} (u)  =  \frac {2 u^{2k-1} K_{2k-1}  (2u)}{\pi \Gamma (2k)},
\tag 67
$$
where $K_{\nu} (x)$ is a modified Bessel function.

For the \qd case, let us consider first the simplest
$\beta= \frac 14 (l=0)$.
In this case we need $G_{\frac 14}(u)$ and  $G_{\frac 34}(u)$.
For $k=\frac 14$, Eq. (66) becomes
$$
\int d(u ; q) u^{2p+1} G_{\frac 14}(u) = \frac 1{ \pi [2]_q}
\prod _{j=1}^p ([j]_{q^2 }  [j- \frac 12]_{q^2})
=\frac { [2p]_q!}{([2]_q)^{2p +1} \pi } \,,
\tag 68
$$
where we use  $[2x]_q = [2]_q [x]_{q^2}$ and
$ [2p]_q! \equiv  \prod_{j=1}^ {2p} [j]_q$.
To solve this, we employ the $q$-exponential function defined as
$$
e_q(\upsilon) \equiv \sum^{\infty}_{n=0} \frac{\upsilon^n}{[n]_q!}, \quad
 \text{for}\,\,\, \upsilon > - \zeta \,\,\, \text{and zero otherwise}
\tag 69
$$
where $-\zeta$ is the largest zero of $e_q(\upsilon) $.
The integration representation of the $q$-factorial is given as
$$
\int^{\infty}_0 d(w;q) e_q(-w) w^m = [m]_q !\,.
\tag 70
$$
{}From this we get
$$
G_{\frac14} (u) = \frac{1}{\pi} \frac{1}{u} e_q (-[2]_q u).
\tag 71
$$
Likewise for $k=\frac 34$, we have
$$
G_{\frac34} (u)
= \frac {[2]_q}{\pi} e_q (-[2]_q u).
\tag 72
$$

The resolution of  unity,  Eq.(65)  is  reduced  to  that  of  the
$q$-coherent states of the $q$-oscillator [19,20] in this case:
$$
\int \frac {d^2(z;q)}{\pi} e_q (-|z|^2) |z)(z| = I.
\tag 73
$$
In fact,  $ |z)$ coincides with the usual
$q$-coherent  states as can be seen in Eqs.(33) and (40),
$$
|z) = \sum^{\infty}_{n=0} \frac{z^n}{\sqrt {[n]_q !}} |n >.
\tag 74
$$
This corresponds to the result of Ref. [14],
`two-component' coherent  states  representations  of  the  $q$-deformed
$osp(1,2)$  superalgebra realized in  terms  of  the  $q$-oscillators
( $|\omega ; k  =  \frac14)$  and  $z  |\omega  ;  k =  \frac34)$ are
denoted as  $||z>_1$ and $||z>_2$  respectively ).

Let us consider case of para-bose oscillator with $l = 1,2,3 \cdots$.
Then we may put $2k-1 = n + \frac12 $ such that $n = 0,1,2, \cdots$
in Eq. (66).
We rewrite  the Eq. (66) as
$$
\int d(u ; q) u^{2p+1} G_k(u) =
\frac 1{ \pi ([2]_q)^{2p+2} [2n+1]_q !}
\cdot [2n+2p+2]_q ! \frac {[p]_{q^2} ! [n]_{q^2} ! }{[n+1+p]_{q^2} !}.
\tag 75
$$
To find $G_k(u)$  we use the integration representation of the
$q$-beta function:
$$
\int^{\infty}_{\alpha^n} d(x ; \alpha) \frac {(x-1)^n_{\alpha}}{x^{n+p+2}}
        =  \frac {[n]_{\alpha} ! [p]_{\alpha}!}{[n+1+p]_\alpha !},
\tag 76
$$
where
$$
(\upsilon + \omega)^n_\alpha \equiv \sum^n_{m=0}
\frac{[n]_\alpha !}{[m]_\alpha ! [n-m]_\alpha !}
        \upsilon^{n-m} \omega^m\,.
\tag 77
$$
Now using the formulae for $q$-factorial and $q$-beta
we have $G_k(u)$ as
$$
G_{\frac{3}{4}+\frac{n}{2}}(u) = \frac{[2]_q}
{\pi^{\frac 32} \prod^n_{j=0} [j+\frac12]_{q^2}} u^{n+\frac12}
        \widetilde K_{n+\frac12}([2]_q u)\,,
\tag 78
$$
where $\tilde K_{n+\frac12} (x)$ is the $q$-deformed modified
Bessel function in integration represention for half-odd integer $\nu$:
$$
\tilde K_{n+\frac12} (x) = (\frac {x}{[2]_q})^{n+\frac12}
                \frac {\sqrt \pi}{[n]_{q^2}!}
        \int^{\infty}_{q^n} d(t;q)  e_q(-xt) (t^2-1)^n_{q^2} \,.
\tag 79
$$

Recalling $\beta  =  \frac14  +  \frac{l}{2}$,  for $l  =  n+1  =
1,2,3,\cdots$
we have  the resolution of unity in Eq. (65) written as
$$
\align
I =&\frac {1}{[2]_q^{n+\frac12}\prod^n_{j=0}[j+\frac12]_{q^2}}
        \int \frac{d^2(z;q)}{2\pi} |z|^{2n+3} \\
         \times &\left[ \{\widetilde K_{n+\frac12} (|z|^2)
        + \tilde K_{n+\frac32} (|z|^2) \} |z)(z| \quad
         + \{\tilde K_{n+\frac12} (|z|^2) -
        \tilde K_{n+\frac32} (|z|^2)\} |z)(-z| \right]\,.
\tag 80
\endalign
$$

Similarly for the case with $l = \frac 12, \frac 32, \frac 52 \cdots$, we
put $2k-1 = n $ where  $n = 0,1,2, \cdots$
in Eq. (66).
We may rewrite  the Eq. (66) as
$$
\int d(u ; q) u^{2p+1} G_k(u) =
 \frac {[p]_{q^2} ! [ p + n]_{q^2} ! }
 {\pi [2]_q [n]_{q^2} !}.
\tag 81
$$
Using the integration representation of the
$q$-factorial in Eq. (70) twice we have $G_k(u)$  as
$$
G_{\frac{n+1}{2}}(u) = \frac{[2]_q}
{\pi [n]_{q^2}!} u^{n}
        \tilde K_{n}([2]_q u)\,,
\tag 82
$$
where $\tilde K_{n} (x)$ is the $q$-deformed modified
Bessel function in integration represention for integer $\nu$:
$$
\tilde K_{n} (x) =\frac {1}{[2]_q}(\frac {x}{[2]_q})^{n}
                \int^{\infty}_{0}  d(t;q^2)
               \frac 1{t^{n+1}}
               e_{q^2}(-t)\,\, e_{q^2}(-\frac{x^2}{([2]_q)^2 t}) \,.
\tag 83
$$
Now the  resolution of unity is written as
$$
\align
I =&\frac {1}{([2]_q)^{n}[n]_{q^2}!}
        \int \frac{d^2(z;q)}{2\pi} |z|^{2n+2} \\
  \times &\left[ \{\tilde K_{n} (|z|^2)
        + \tilde K_{n+1} (|z|^2) \} |z)(z| \quad
         + \{\tilde K_{n} (|z|^2) -
        \tilde K_{n+1} (|z|^2)\} |z)(-z| \right]\,.
\tag 84
\endalign
$$

We may normalize the $q$-coherent states $|\omega ; k\! >$ as
$$
|\omega ; k\! > =\frac {|\omega ; k)} {  N_k(|\omega|^2 )}\,,
\tag 85
$$
where using Eqs.(49) and (38),
$$
N_k^2 (|\omega|^2 )
= (\omega ; k | \omega ; k) =
        \sum^{\infty}_{p=0}\,\, \frac {|\omega|^{2p}}{[p]_{q^2} !
        \prod_{j=1}^p[j+2k-1]_{q^2}!},
\tag 86
$$
which is identified as a $q$-hypergeometric function.  Note also that
$$
\align
&<\omega';\beta |\omega ; \beta + \frac12>= 0, \\
&<\omega'; k |\omega ; k>
= \frac {N_k^2 (\omega'^* \omega )}{ N_k(|\omega'|^2) N_k (|\omega|^2 ) }\,,
\tag 87
\endalign
$$
showing that the $q$-coherent states are not orthogonal in each
$D_{\beta}$ or $D_{\beta+ \frac12}$.
Using the above Eqs.(85)-(87) and (55)-(56), the normalized
$|z>$ is given as
$$
\align
&|z > = \frac{  |z)} {{\Cal N}(|z|^2)} ,\\
&{\Cal N}^2 (|z|^2) = (z|z)
= N_{\beta}^2 (|\omega|^2  ) + \frac{|z|^2}{[4\beta]_q}
                N_{\beta + \frac12}^2 (|\omega|^2  ), \\
 & <z' |z> = \frac { {\Cal N}^2 (z'^{*} z)}
                    {{\Cal N}(|z'|^2) {\Cal N} (|z|^2)} \,.
\tag 88
\endalign
$$

\bigskip
\centerline {\bf 4. Summary and Remarks}
\bigskip

We have found a $q$-deformed version of para-bose oscillator
associated with two-body Calogero model.
It realizes the $su_{q^2} (1,1)$ algebra whose form looks same
as in the standard $q$-oscillator case.
Especially we note that the $q$-oscillator
$A_q$ and $A_q^+$ are invariant under $q \to q^{-1}$
in contrast with the ones considered  in Ref. [12], while they share the
same Fock space.

The  $q$-coherent  states  of the para-bose oscillators are
also constructed
and the resolution of unity
is demonstrated for order $2l +1= 1, 2, 3\cdots$
employing the integration representation of
\qd exponential and \qd beta function.
For non-integral value of $2l+1$, we note that the proof for
the resolution of unity may need carefully developed
$q$-special functions.

Also we remark that for more than two-body Calogero model,
the $q$-deformation does not go parallel with that of two-body case.
The method given in the text is not generalized straight-forwardly
to the many-body case.
In addition, the equivalent oscillators for
$N (\ge 3)$-body Calogero model [5]  do not satisfy the para-bose algebra.
Therefore, the $q$-deformation for the many-body case
seems quite a  challenging problem.

\bigskip
\centerline {\bf Acknowledgement}

The authors wish to thank Prof. K. H. Lee for useful  discussions
about solving difference equations. This work is supported in part by
KOSEF 931-0200-030-2 and through CTP in SNU{}.

\vfill\eject

{\bf References}
\bigskip

\item{[1]} Drinfeld V G 1986 {\it Proc. ICM } (Berkeley) 798
    \item{} Jimbo M 1985 {\it Lett. Math. Phys.} {\bf 10} 63
\item {[2]}  Kulish P P and Damaskinsky E V 1990
            {\it J. Phys. A : Math. Gen.}  {\bf 23}  L415
     \item{}  Ui H and Aizawa N 1990 {\it Mod. Phys. Lett. A}
            {\bf 5}  237
\item {[3]} Floreanini R and Vinet L 1990 {\it J. Phys. A: Math. Gen.}
            {\bf 23} L1019
     \item {} Odaka K, Kishi T and Kamefuchi S 1991 {\it J. Phys. A: Math.
            Gen.} {\bf 24} L591
\item{[4]} Floreanini R, Spiridonov V P and Vinet L 1990
          {\it Phys. Lett. } {\bf B242} 383
      \item{} Palev T D 1993 {\it Quantization of }
            $U_q [osp(1/2N)]$ {\it with   Deformed Para-Bose Opeators}
             ICTP preprint
\item {[5]}  Polychronakos A P 1992 {\it Phys. Rev. Lett.}
             {\bf 69} 703
       \item{}  Brink L, Hansson T H and Vasiliev M A 1992
              {\it  Phys.  Lett. B}  {\bf 286} 109
       \item{} Brink L, Hansson T H, Konstein S and Vasiliev M A 1993
        {\it Nucl. Phys. B} {\bf 401} 591
\item {[6]}  Calogero F 1971 {\it J. Math. Phys.} {\bf 12} 419
    \item{}  Gambardella P J 1975 {\it J. Math. Phys.} {\bf 16} 1172
    \item{}  Olshanetsky M A and Perelomov A M 1983 {\it Phys. Rep.}
             {\bf 94} 313
\item {[7]}  Wybourne B G 1974 {\it  Classical  Groups  for  Physicists}
            (New  York :  Wiley)
\item {[8]} Green H S 1953 {\it Phys. Rev. } {\bf 90} 270
       \item{} Ohnuki Y and Kamefuchi S 1982 {\it Quantum Field
             Theory and Parastatistics}  (Tokyo: University of Tokyo
             Press; Berlin Springer)
\item {[9]}    Mukunda N, Sudarshan E C G, Sharma J K and Mehta C L 1980
                {\it J. Math. Phys.}{\bf 21} 2386
        \item {} Ohnuki Y and Kamefuchi S 1978 {\it J. Math. Phys.} {\bf 19} 67
\item{[10]} Ganchev A Ch and Palev T D 1980 {\it J. Math. Phys.}
          {\bf 21} 797
\item {[11]}  Macfarlane A J 1989 {\it J. Phys. A : Math. Gen.}
             {\bf 22} 4581
       \item{}  Biedenharn L C 1989 {\it J. Phys. A : Math. Gen.}
             {\bf 22} L873
\item {[12]}    Brzezinski T, Egusquiza I L and Macfarlane A J 1993
                {\it  Phys.  Lett.} B{\bf 311} 202
\item {[13]}    Sharma J K, Mehta C L, Mukunda N and Sudarshan E C G 1981
                {\it J. Math. Phys.}{\bf 22} 78
\item {[14]}    Kuang L-M, Zeng G-J and Wang F-B 1993
{\it J.  Phys.  A  ;  Math. 
 Gen.}
 {\bf 26} 4011
\item {[15]}    Cutright T L and Zachos C K 1990 {\it Phys.  Lett.}
B{\bf  243} 237
\item {[16]}    Barut A O and Girardello L 1971 {\it Commun. Math. Phys.}
                {\bf 21} 41
\item {[17]}    Chaichian M, Ellinas D and Kulish P 1990
{\it Phys. Rev. Lett.} {\bf 65}, 980
     \item{}    Ellinas D 1993 {\it J. Phys. A : Math. Gen.} {\bf 26} L543
     \item{}    Quesne C 1991 {\it Phys. Lett. A}{\bf 153} 303
     \item{}    Gong R 1992 {\it J. Phys. A : Math. Gen.}\,\, {\bf 25} L1145
\item {[18]}    Floreanini R and Vinet L 1992 {\it Phys. Lett.}
{\bf B 277} 442
\item {[19]}    Gray R W and Nelson C A 1990 {\it J. Phys. A : Math. Gen.}
                {\bf 23} L945
\item {[20]}    Bracken A J, McAnally D S, Zhang R B and
                Gould M D 1991 {\it J. Phys. A : Math. Gen.} {\bf 24} 1379

\vfill
\bye